# High Speed Precise Refractive Index Modification for Photonic Chips through Phase Aberrated Pulsed Lasers

Bangshan Sun[1*], Simon Moser[2], Alexander Jesacher[2*], Patrick S. Salter[1], Robert R. Thomson[3], Martin J. Booth[1*]

[1]Department of Engineering Science, University of Oxford, Oxford OX1 3PJ, United Kingdom
[2]Institute of Biomedical Physics, Medical University of Innsbruck, Müllerstraße 44, 6020 Innsbruck, Austria
[3]Institute of Photonics and Quantum Sciences, Heriot-Watt University, Edinburgh EH14 4AS, UK

*Corresponding to: Bangshan Sun (bangshan.sun@eng.ox.ac.uk), or Alexander Jesacher (alexander.jesacher@i-med.ac.at), or Martin J. Booth (martin.booth@eng.ox.ac.uk)

## Abstract

Integrated photonic chips have significant potential in telecommunications, classic computing, quantum systems, and topological photonics. Direct laser writing offers unique capability for creating three-dimensional photonic devices in an optical glass chip with quick prototyping. However, existing laser writing schemes cannot create index-modified structures in glass that precisely match the laser focal shape while also achieving high scanning speed and high refractive index contrast. Here, we introduce the theory of a refractive index modification scheme that combines the advantages of both traditional non-thermal and thermal regime fabrication methods. We also propose a model of waveguide formation that was verified through a thorough study on the effects of phase aberrations on the laser focus. The presented new photonic chip fabrication scheme uses a novel focal intensity distribution, where pulse energy is relocated to the bottom of a laser focus by manipulating primary and higher order spherical aberrations. The technique can produce index modifications with high scanning speed (20 mm/s or higher), high index contrast (16 × $10^{-3}$), and high precision to fabricate with arbitrary cross-sections. This method has potential to expand the capabilities of photonic chips in applications that require small-scale, high precision, or high contrast refractive index control.

## 1. Introduction

Integrated photonic chips[1,2] play a crucial role in many applications, thus enhancements in related technologies are important and could lead to significant impacts. Optical glass is commonly used in integrated photonic chips, owing to its transparency across a wide range of wavelengths, spanning from ultra-violet to visible and near infrared. High-precision refractive index (RI) modification is essential for the creation of fundamental elements such as micro-waveguide based photonic devices[3] in most integrated chips, especially in applications such as telecommunications[4,5], quantum systems[6–8], and topological photonics[9–12].

Multiple techniques can create refractive index modification for fabricating on-chip photonic devices. One of the most prevalent methods is semiconductor fabrication technology[13,14], which involves deposition, optical lithography, and etching. This approach is highly effective in creating low-loss waveguides on a large scale, but it has limitations such as the inability to create three-dimensional (3D) devices and lengthy prototyping times. Another approach is femtosecond laser (fs-laser) direct writing[4,5,15], which is a promising method for creating on-chip devices by index modification in three dimensions and with rapid prototyping capabilities[16]. Most recently, laser-written waveguides have found significant potential in the development of integrated photonic chips for quantum[6,17–22] and topological photonic research[9–12].

Currently, there are two distinct regimes[4,5] employed for fs-laser written waveguides in optical glass: a non-thermal regime utilizing low laser repetition rates (< 100 kHz), and a thermal regime utilizing high laser repetition rates (> 500 kHz). These two regimes induce different RI modifications with different physical dimensions. In the non-thermal regime, it is possible to stack multiple scans of laser spots closely together to create structures with precisely controlled cross-sections. However, a significant limitation is that this requires very slow scanning speeds (0.01 - 0.1 mm/s)[4,5] to achieve continuous index modification. Functional device fabrication may take several hours, this being susceptible to disturbance. Another limitation of non-thermal regime fabrication is that the refractive index contrast achieved is typically low[5] ($< 3 \times 10^{-3}$). On the other hand, in the thermal regime fabrication, the efficiency can be significantly increased, reaching scanning speeds of 2 - 15 mm/s (2 to 3 orders of magnitude higher), and with higher RI contrast (e.g., $4 - 8 \times 10^{-3}$ in borosilicate glass). However, the upper limit of RI contrast is still not sufficient for certain applications, and there is not an efficient way to produce accurate high contrast refractive index profile with high fabrication precision. Another major drawback of thermal regime fabrication is that the modified index structure is unrelated to the laser focal shape and is typically much larger, making it very difficult to create precise small scale photonic devices. This is also revealed in the multiscan scheme[23–25], where features are built by scanning the laser spot multiple times with spatial shift. The precision of thermal regime multiscan is limited along the axial direction of laser propagation, making it challenging to stack scans with close distances (e.g., < 8 μm) axially, thereby significantly limiting the precision of functional on-chip devices.

In this work, we present the underlying mechanism of a high speed and high precision refractive index modification scheme in optical glass, which overcomes several key fabrication limitations of existing direct laser writing methods. We conduct a systematic investigation into the effects of phase aberrations on refractive index modification in optical glass. We then propose a hypothesis model of glass waveguide formation, stemming from the impact of heat accumulation and sequence of regional index modification resulting from different laser focal shapes. The details of the fabrication schemes are described thoroughly by examining single scan and multiscan techniques, as well as by constructing target 3D energy relocated focal intensity distribution using a phase retrieval algorithm that combines multiple Zernike phase modes, including primary and higher orders spherical aberrations. Our method demonstrates a rapid and precise laser writing technique for creating 3D functional on-chip photonic devices in optical glass with high index contrast and low loss.

## 2. Results and Discussion

### 2.1. Effect of Zernike Phase Aberrations & Waveguide Formation Model

We conducted a comprehensive study on the effects of phase aberrations on refractive index modification in optical glass, as there is currently limited literature on this topic. The study was conducted in thermal regime, as we aimed to achieve high scanning speed for index modifications. Zernike polynomials are commonly used to represent phase aberrations in optical systems. We investigated their effects using simulations of focal intensity based on Fourier optics[26], as well as RI profile images induced by ultrafast lasers with phase aberrations controlled using a spatial light modulator (SLM). We applied individually common Zernike modes in Noll's sequential indices with an RMS amplitude of -1 rad, including Z5 (astigmatism), Z6 (astigmatism), Z7 (coma), Z8 (coma), Z9 (trefoil), Z10 (trefoil), and Z11 (spherical) to the optical system. The details of these Zernike aberrations are summarised in Supplementary Information Table S1. The simulation of axial intensity profile of the laser focus with these aberrations is shown in Figure 1 (b). To obtain a more accurate evaluation and clearer picture of the laser-induced heat accumulation inside the glass, we also calculated the integrated focal intensity by integrating the light intensity along the third axis (y axis, as shown in supplementary information Figure S1 and Note 1).

We specifically observe that when Z11 (spherical) = -1 (last column in Figure 1), it represents the inverse scenario compared to the diffraction caused by refractive index mismatch[27]. In this case, marginal rays converge more closely to the sample surface, forming a small focal point accompanied by diffraction rings. Paraxial rays converge farther from the interface. Due to their smaller numerical aperture (NA), the resulting focus is larger, characterized by the presence of a major lobe in the laser focus. Cross-section images of the aberrated laser-induced waveguides are presented in Figure 1 (d) and (e), obtained using two different pulse energies from the laser, resulting in different levels of heat accumulation.

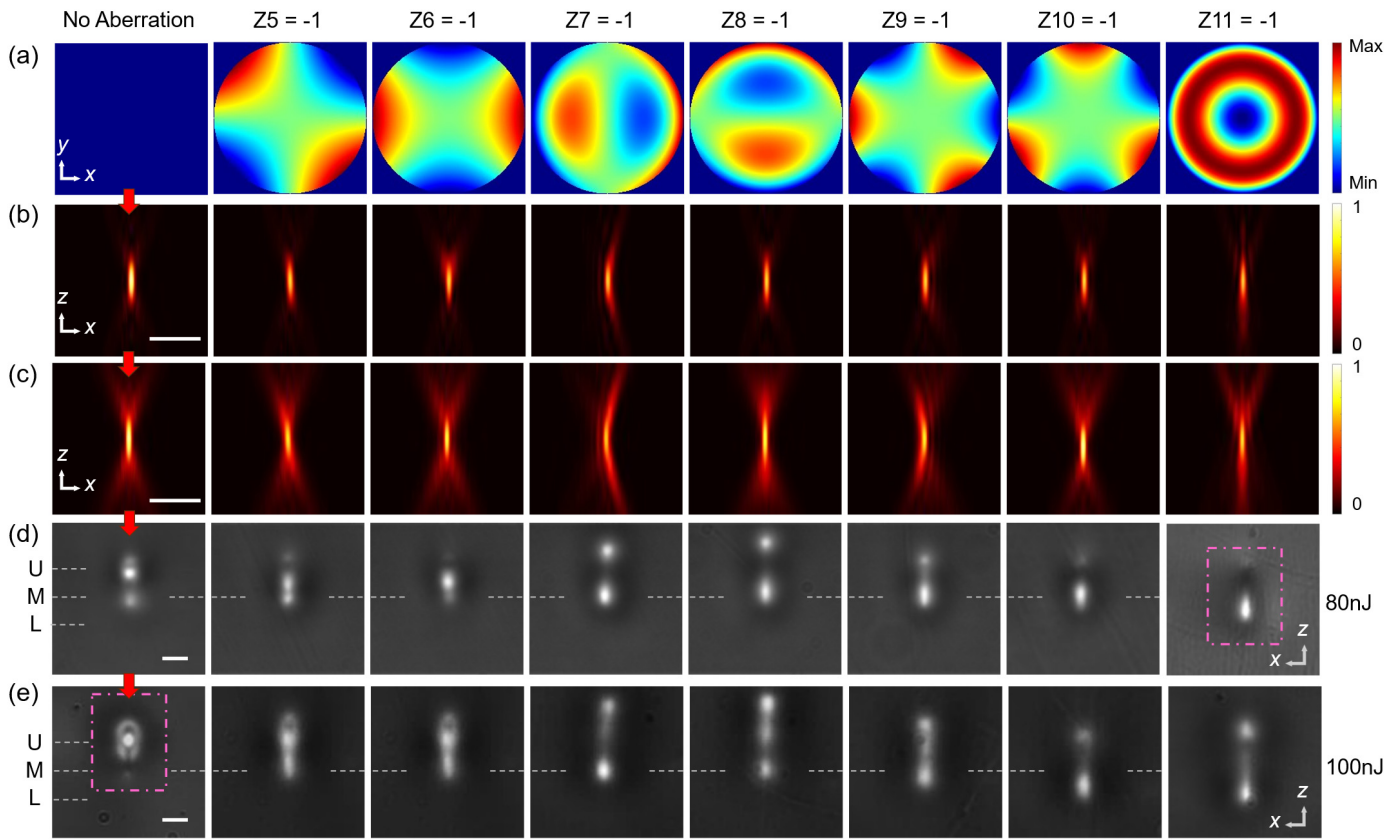

**Figure 1**. Effect of Zernike phase aberrations on refractive index modification in borosilicate glass (Corning EAGLE 2000). Optical system and 3D spatial x-y-z axes are shown in Supplementary Information Figure S1. Red arrows in first column mark the laser propagation direction. (a) Plot of Zernike phase aberrations at the objective pupil. (b) Simulated axial intensity profile of aberrated laser focus at the position y = 0. All simulation images were normalized to the peak intensity of the case without aberration (left image). (c) Simulated accumulated axial intensity profile of aberrated laser focus in the range of two Airy units (Supplementary Note 1). All simulation images were normalized to the peak accumulated intensity of the case without aberration (left image). The sidelobes have little effect on the glass index modification due to low accumulated heat. (d)-(e) Broadband visible light microscopic images of waveguide cross-sections formed in the presence of phase aberrations. Dashed straight lines denote three spatial position relative to laser focus: U represents upper region, M represents middle region, and L represents lower region. Laser pulse energy measured at the objective pupil was 80 nJ and 100 nJ for (d) and (e) respectively. The dashed lines mark a fixed spatial position to compare the axial shifts of the waveguides. As spherical aberration (Zernike mode 11) led to a larger axial shift (~10 μm in this case), we adjusted features in both simulations and experiments to be centred to the images. Dashed line rectangular boxes mark two high quality waveguide formation regimes. Scale bars: 5 μm. Note, to show more laser focal details, the field of view of the simulation images in (b)(c) is half of that in (d)(e).

These results allowed us to hypothesise a model to predict refractive index modifications based on different focal shapes. This model, which is summarised in Figure 2, was based on the accumulated heat distribution, sequence of index modifications across different regions around laser focus and its impact on laser propagation and focusing. We first analysed the case of a laser focus without any aberration, as shown in the left two images in Figure 1 (d) and (e). Its corresponding model is presented in the first column of Figure 2. We observed that, when fabricated with lower pulse energy (Figure 1 (d)), the resulting waveguide cross-section had two lobes, despite the focus being diffraction limited (Figure 1 (b)(c)). We hypothesise two factors contributing to this phenomenon. The first, is that heat accumulation is biased towards the upper part of the focal region, because pulses get absorbed earlier while the plasma starts to form. The second, which could be the major factor, is that when the laser pulse propagates through the focus, the refractive index in the upper region becomes modified first. This modification, together with the high-density plasma concentrated in the upper region affect the laser propagation and normal focusing, thus disrupting the tight focusing property at the middle and bottom parts of the focus. As such, the middle part of the focus experiences reduced intensity, but it is still strong enough to accumulate heat for RI modification. The region between top and middle parts, as well as the bottom part, accumulate insufficient heat for index modification, leading to the formation of two-lobe waveguides at the upper and middle parts, respectively. As we increased the input laser power, the index modification at the upper part became stronger and larger, causing more significant disruption of the focus in the centre. In such case, the second lobe became less significant with increasing laser power, as shown in Figure 1 (e). This behaviour is summarized as first column in the model presented in Figure 2. Interestingly, this also explains why it is challenging to reduce the lateral size of thermal regime fabricated waveguides by simply reducing laser power, as doing so may result in a multiple-lobed structures.

Next, we analysed the case of a focus significantly elongated along the axial z direction, represented by Z7, Z8, and Z9 in Figure 1. The middle column in Figure 2 summarizes the waveguide formation model for this case. When fabricated with lower laser power, the phenomenon of two lobes can be explained in a similar manner as above. However, when fabricated with higher laser power, compared to the normal focus case, the RI modification at the upper focal region has limited negative effect on the laser focusing at the middle region due to elongated focus, which is why the middle region lobe did not disappear in this case.

The third case we analysed was a focal shape with the laser energy concentrated more towards the bottom of the focus ("energy relocated focus"), represented by $Z11 = -1$ in the last column of Figure 1, with corresponding model at last column in Figure 2. It is notable that Z11 creates surrounding sidelobes on the upper part of the focus, as shown in the last image of Figure 1 (c). These sidelobes were relatively distant from the centre and had low intensity, making it challenging to achieve effective heat accumulation. Our experiments revealed no discernible material changes resulting from these sidelobes. As shown in Figure 2 (a), this type of intensity distribution created more obstacles for heat accumulation at the upper part of the focus, meaning that the RI modifications could occur at the middle and bottom parts before they were heavily affected by the upper part. With lower laser power, fabrication with this type of focal intensity distribution could produce a modified index structure that nearly matches the laser focal shape and size, as demonstrated in the right image of Figure 1 (d). When increasing the input laser power, the upper part region could form a level of index modification that affected laser propagation, producing a two-lobe structure similar to the other cases.

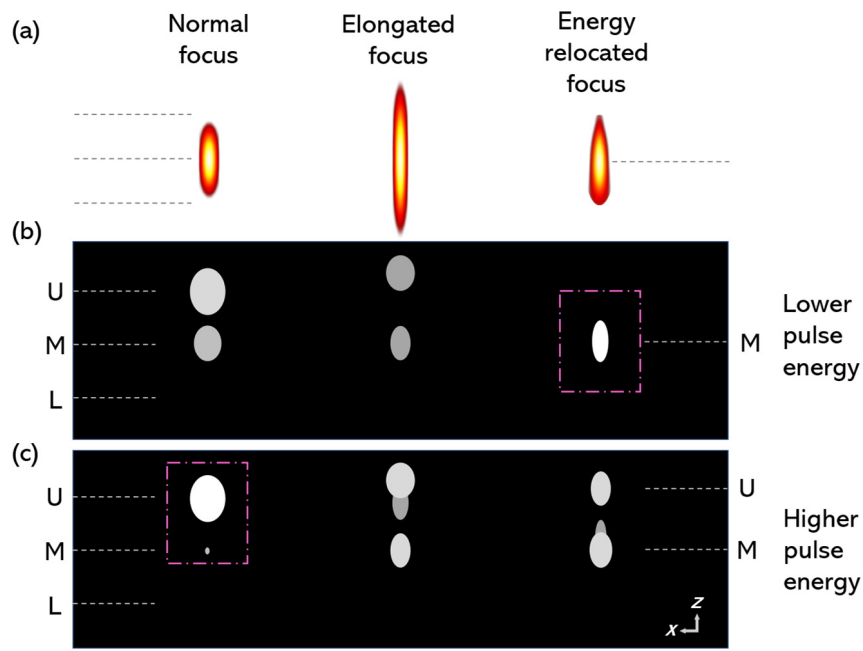

**Figure 2**. Hypothesised models for waveguide formation in borosilicate glass. Dashed straight lines denote three spatial position relative to laser focus: U represents upper region, M represents middle region, and L represents lower region. (a) Diagram depicting three common focal shapes. (b)-(c) Diagrams illustrating the cross-sectional shapes of the formed refractive index modifications corresponding to the ultrafast laser focal shapes shown in (a). Varying levels of brightness indicate the magnitude of refractive index modification. (b) Lower input pulse energy. Dashed box highlights the new formation regime characterized by an energy-relocated focus. (c) Higher input pulse energy. Dashed box highlights the conventional thermal regime waveguide formation with a normal focus.

The two valid regimes for waveguide fabrication are highlighted with dashed boxes in Figure 1 and Figure 2. The regime created by a normal focus represents the traditional thermal region fabrication, while the regime created by an energy-relocated focus represents a new index modification region ideal for high-precision photonic chips. To validate the model depicted in Figure 2, we conducted direct laser writing experiments with varying laser conditions. These experiments were carried out at different sample depths, scanning speeds, laser power settings, in at least 40 individual borosilicate glass photonic chips over a three-year timeframe. The obtained results were highly repeatable, and the model applies to nearly all the obtained results.

## 2.2 Index Modification with Energy Relocated Focus

To investigate the effects of focal intensity relocation on refractive index modification, we conducted an extensive study by varying the amplitude of Z11. We compared the simulation of focal intensity distributions with widefield microscopic images of cross-sections. We also imaged the refractive index profiles and laser guiding mode profiles for detailed comparison, as shown in Figure 3. When positive Z11 values were applied, the focal intensity distribution exhibited patterns similar to "energy relocated focus" (shown in Figure 2(a)), but inverted along the z direction, resulting in poor waveguide formation. As the Z11 values decreased from 0, the focal energy was increasingly relocated towards the bottom part. Comparing the index modification structures for different values of Z11, such as Z11 = 0, -0.3, -0.6, and -0.9, we observed that the guiding of the top lobes became weaker, while the guiding of the bottom part became stronger (Figure 3 (c)). When the amplitude of Z11 exceeded a certain threshold (around -0.8 in this case), the top guiding lobe nearly disappeared, and the index modification structure of the bottom lobe matched the focal intensity distribution well. Further increase in the negative amplitude of Z11 resulted in more focal distortion, causing a weaker laser focus and deteriorated RI modification. We found experimentally that the optimal amplitude of Z11 for achieving precise index modification was around -1, with a recommended range of -0.8 to -1.5, for an objective lens with NA of 0.5. This optimal amplitude for Z11 however varies with the NA of objective but can be easily explored by experiment. Comparison of the RI profiles of Z11 = 0 and Z11 = -1 revealed that Z11 = -1 resulted in much higher RI contrast and smaller lateral size, which are crucial properties for high-precision photonic devices.

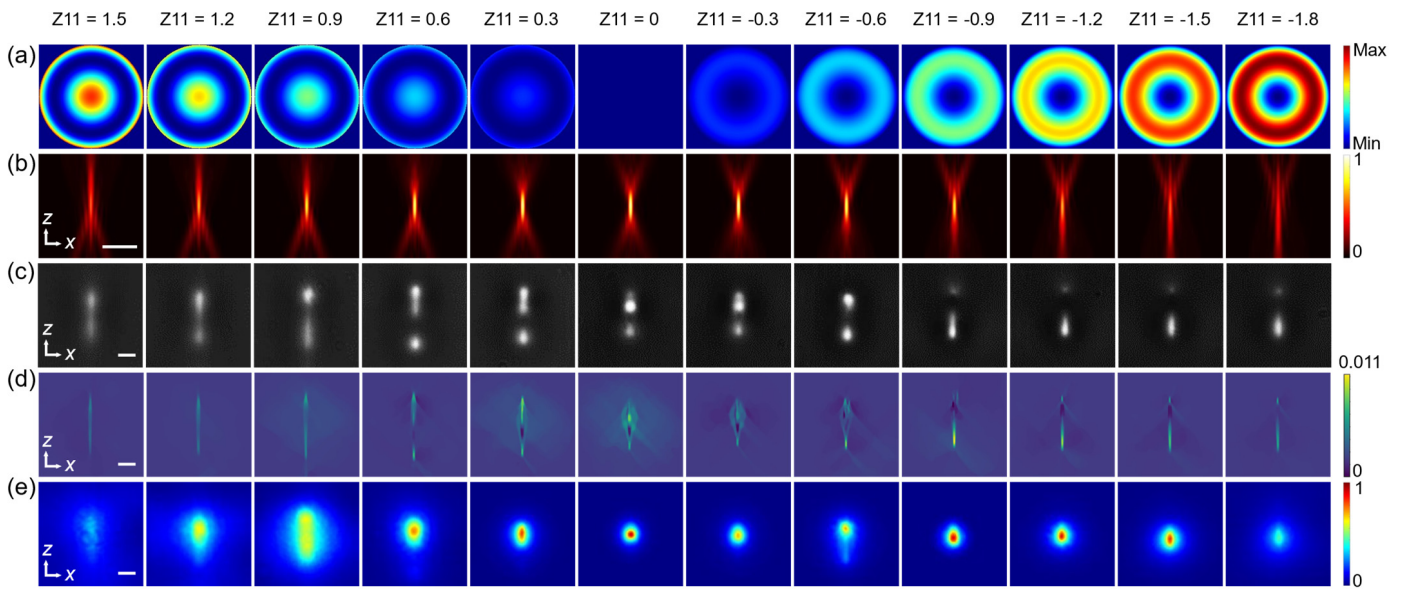

**Figure 3**. The new high-performing index modification regime arises by relocating more laser energy to the bottom of a focus. Zernike mode 11 is adapted to control the focal energy distribution. The objective NA was 0.5. (a) Plot of Zernike mode 11 with varying amplitude at objective pupil. (b) Accumulated axial intensity profile of laser focus (Supplementary Note 1). The sidelobes had little effect on the index modification. All simulation images were normalized to the peak accumulated intensity of the case without aberration (Z11 = 0). (c) Broadband visible light microscopic images of corresponding waveguides cross-sections. (d) 3D tomographic microscopic refractive index images of waveguide cross-sections. Enlarged version in Supplementary Figure S3. (e) 785 nm laser guiding mode profiles. The new fabrication regime emerged when laser energy was sufficiently relocated to the bottom. Scale bars: 5 μm. Laser pulse energy was 80 nJ. Laser scanning speed was 8 mm/s.

In most cases, traditional fs-laser direct writing with Z11 = 0 is adequate for fabricating waveguide photonic chips with single scan or large dimensions[4]. However, there are certain applications that require photonic devices with smaller scales, such as non-circular structures with dimensions less than 10 μm. Fabricating such structures using traditional thermal regime fabrication is extremely challenging due to its limited precision[3]. On the other hand, using traditional non-thermal regime fabrication method results in low RI contrast structures and long fabrication times with relative higher failure rates. In contrast, we demonstrated the unique capabilities of this new fabrication scheme in creating high-precision, high RI contrast structure for low-loss photonic chips in glass.

To create non-circular structures, multiscan fabrication is typically required[23–25]. However, in the high-speed thermal region fabrication regime, it is hard to precisely control the cross-section shape. Especially, precisely controlled RI modification in small sizes (e.g., < 8μm) is very challenging[24]. As depicted in Figure 4, using traditional thermal regimes to fabricate small-scale non-circular structures is challenging because when the multiscan laser spots are too close together, the new scans may erase existing high index areas. This may be because fs-laser writing in glass can induce both positive and negative index regions. In contrast, the energy relocated fabrication scheme with Z11 = -1 allows closely stacked multiscan without any issues, as the negative index structure in a single scan is only in the upper part. Consecutive scans overlap perfectly with existing structures, bringing much improved fabrication precision.

Another significant advantage of the energy relocated fabrication scheme is the capability to create structures with much higher RI contrast, as evident in the refractive index profiles shown in Figure 4. Comparing the laser guiding mode profiles, structures fabricated using traditional thermal regimes exhibit poor mode profiles, which are clearly distorted, and highly lossy waveguides. On the other hand, the energy relocated fabrication scheme produces improved mode profiles, leading to much better quality of laser guidance.

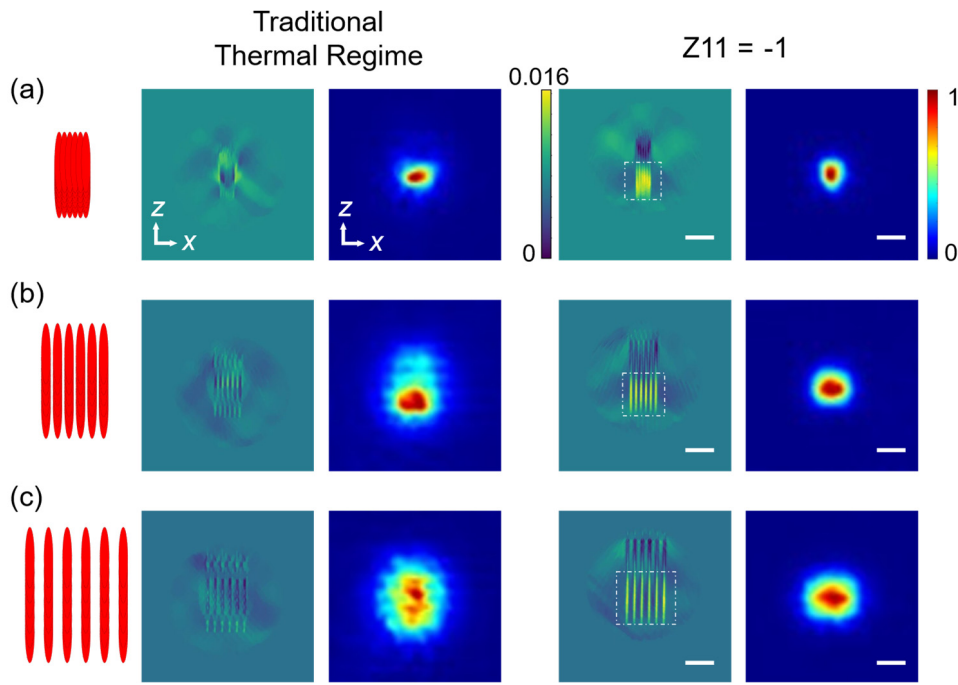

**Figure 4**. High-precision fabrication of small-scale photonic devices. All fabrication was done with 6×6 scans along x and z. (a)-(c) Multiscan fabrication with lateral and axial separations of 0.5 μm, 1.0 μm, 1.5 μm respectively. Traditional thermal regime fabrication with Z11 = 0 was compared with the energy relocated fabrication scheme with Z11 = -1. The red sketch on the left highlights the multiscan. Refractive index profiles were compared with 1550 nm laser guiding profiles. Enlarged version of RI profiles are in Supplementary Figure S4. Scale bars: 5 μm. Laser pulse energy was 100 nJ for the traditional thermal regime fabrication, and 80 nJ for the energy relocated scheme with Z11 = -1. Laser scanning speed was 8 mm/s.

**2.3 Algorithm for Higher Order Spherical Phases**

Our ultimate target is to produce the relocated focal intensity that has pulse energy shifted to the bottom of the laser focus (as shown as "Energy relocated focus" in Figure 2). While applying a Z11 phase to the input laser light can generate a focus close to the target intensity distribution, it is unable to accurately reproduce such a distribution. This is evident by a number of side lobes in the simulated focal intensity profiles of Z11 = -1 (Figure 1 (b)(c)). While the side lobes away from the centre (x = 0) have no noticeable effect on the index modification, the top lobe, which is right above the peak intensity region, has considerable negative impact, mainly stretching the modified index region along the z direction. This is unfavourable in applications where small size (< 3μm) refractive index modifications are needed.

To address this problem, and to implement the target ideal "energy relocated focus" shown in Figure 2, we developed an algorithm using the Nelder-Mead method to combine a number of Zernike modes (Figure 5 (a)). The first step was to create a 3D focal intensity distribution, following the "energy relocated focus" case depicted in Figure 2, as our target. During each iteration, the algorithm simulated a 3D focal intensity based on a guess of input phase, calculated the mean-squared-error (MSE) between the simulated and target 3D foci, and used the MSE as feedback to update the input phase. We set a target MSE tolerance to ensure algorithm convergence. Since the target focal intensity distribution may follow standard spherical phases, we included in the algorithm primary and higher-order spherical phases, including Z11, Z22, Z37, and Z56. Details of these spherical aberrations are included in Supplementary Information Table S1. The results suggested that a combination of a major contribution from Z11 with an amplitude of -1, a minor contribution from Z37 with an amplitude of -0.3, and negligible contributions from Z22 and Z56.

The demonstration results are summarized in Figure 5, where Z11 = -1 was applied in all cases. To demonstrate effectiveness, we compared the simulated focal intensity with and without higher-order spherical aberration of Z37. The top left simulated intensity images clearly show that after application of Z37, the sidelobes, and especially the important top lobe, were suppressed. To highlight the change in focal shape, we plotted the normalized intensity (summed along x) versus axial z coordinate in the right figure of Figure 5 (b). A Gaussian distribution was added for reference. It is evident that when applying only primary spherical aberration of Z11 (red curve), there were obvious lobes on top and bottom. After Z37 was applied with an amplitude of -0.3 (blue curve), these lobes were significantly suppressed, while the overall intensity distribution still had energy relocated to the bottom (as compared to the dashed Gaussian curve). The imaged refractive index profiles (bottom left images in Figure 5 (b)) indicate a reduced extension of the high index region along the axial z direction after Z37 was applied, which represents a considerable improvement.

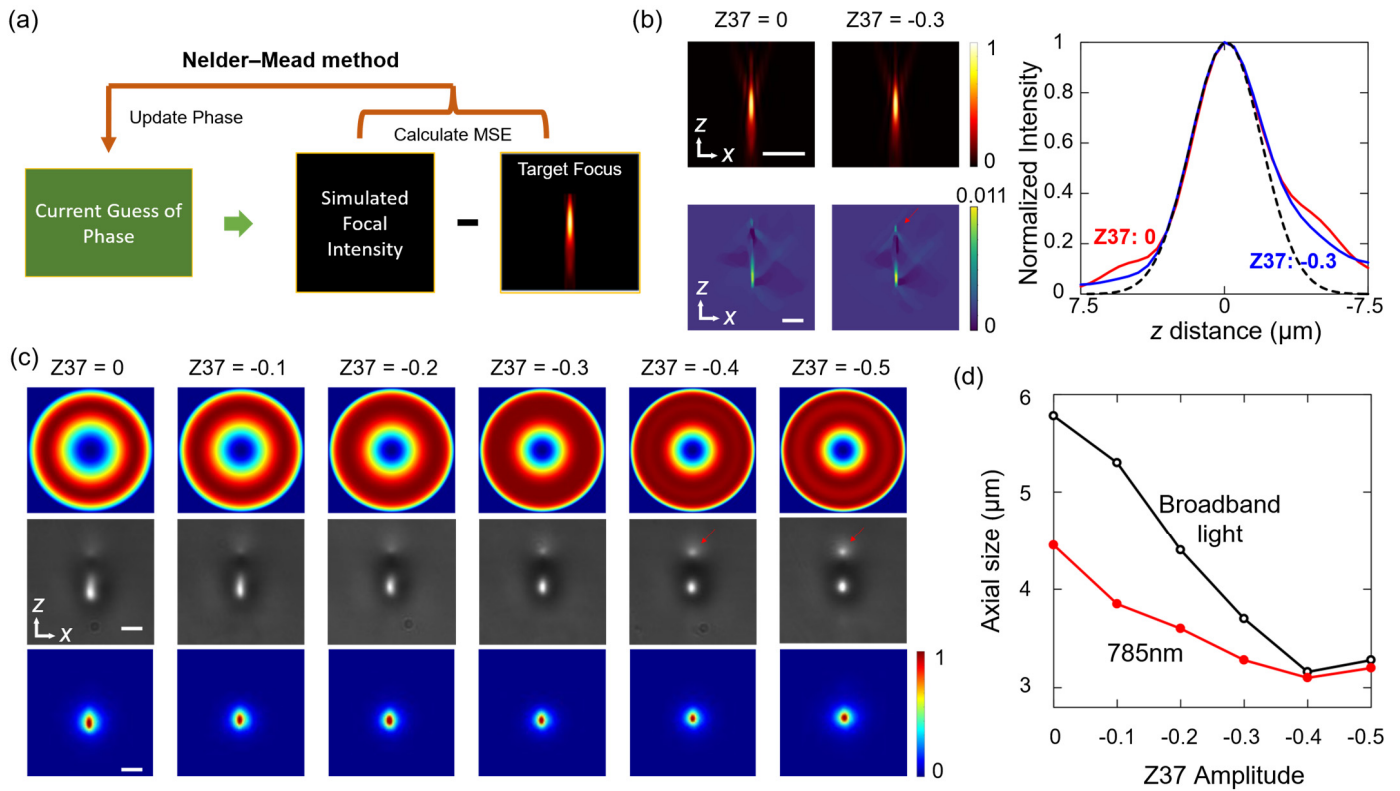

**Figure 5**. Results of the Nelder-Mead optimization, which suggested combining primary and higher order spherical phases to achieve much improved energy-relocated focal intensity distribution and dimension control of a modified structure. In all cases, a base phase of Z11 = -1 was applied to the input laser, while the amplitudes of additional Z37 are stated in the figure. (a) Flow chart of the Nelder-Mead method to optimize the phase. (b) Comparing the case without Z37 to the case with an additional amplitude of Z37 = -0.3. Top left: simulated focal intensity profiles. Bottom left: corresponding measured refractive index profiles (enlarged version in Supplementary Figure S5). Right: plot of the normalized simulated focal intensity along the axial z direction. This plot compares the results of no additional Z37 (red curve), additional Z37 = -0.3 (blue curve), and a Gaussian distribution (black dashed curve). (c) Dimension control by additional Z37. Top: Introduced phase combining Zernike mode 11 with amplitude of -1 and Zernike mode Z37 with varying amplitudes. Middle: broadband widefield microscope images of single scan cross-sections for varying additional Z37 applied. Bottom: 785 nm laser guiding mode profiles. (d) The measured axial sizes (diameter between 1/e peak intensities) versus Z37 amplitudes for broadband widefield microscopic images (black) and 785 nm laser guiding mode profiles (red). Scale bars: 5 μm. Laser scanning speed was 8 mm/s.

It is important to reduce the axial size of high index structure produced by the single scan of this new scheme, as this would further reduce the minimal size of any photonic device along the axial z direction. We demonstrated the effect of this axial size suppression by gradually increasing the negative amplitude of Z37. The cross-section images of these laser processed structures are shown in Figure 5 (c), and their axial sizes are plotted in Figure 5 (d). It is evident that the axial size of cross-section from broadband widefield microscopic images is considerably reduced from 5.8 μm to 3.1 μm after adding Z37 from 0 to -0.4, the axial size of corresponding laser guiding mode of 785 nm laser is also reduced from 4.4 μm to 3.1 μm.

It is noticeable that there is a small high refractive index region at the top (bottom index profiles in Figure 5 (b)), which was generated by the strain of the negative index region in the middle region. Further increasing the negative amplitude of Z37 beyond -0.3 would increase this small high index at the top, shown as the LED widefield microscopic images in Figure 5 (c). This can also be understood as bigger amplitude of Z37 makes the focal intensity to be closer to a laser focus without aberration (Gaussian-like), which generates two lobes as shown as the left image in Figure 1 (d). We also noticed in Figure 5 (d) that the effect of reducing axial size becomes weaker when applied Z37 is beyond -0.3. Considering all these, applying Z37 with amplitude of -0.3 to -0.4 would be the optimal choice for an objective with NA of 0.5. This optimal amplitude is of course related to objective NA and can be explored for individual objective lens. In practical applications, Z37 is crucial for creating photonic devices which require a small scale along the z direction (for example, < 3 μm). Using a higher NA objective lens would also be able to further reduce this size.

## 2.4 Discussion

One advantage of the presented index modification scheme is its ability to closely stack multiple scans with various arrangements, even with high scanning speed in the thermal fabrication regime, while maintaining the integrity of the existing fabricated structure. This advantage facilitates the creation of 3D photonic devices with virtually any desired cross-section, ensuring high precision, and enabling fast fabrication speeds. Such capabilities are challenging to attain through conventional methods. During our investigations, high quality photonic devices were fabricated by adopting varying laser scanning speed ranging from 2 to 20 mm/s.

An additional distinctive advantage lies in the capability to achieve high contrast and controllable RI modifications. Through our experimental investigations, we found that by multiscan without any spatial shift over existing index modified structure, we were able to precise control the peak RI contrast within the range of 0.009 to 0.021 by simply adjusting the number of multi-scans (e.g., a peak RI contrast of 0.021 was obtained by 12 scans over the same modified structure). This property enables applications requiring small bending radiuses or low bending losses. Moreover, the increased RI contrast facilitates the possibility of superior confinement of guided light, effectively minimizing cross-coupling between adjacent waveguides. This characteristic positions this technique as an excellent choice for developing integrated photonic chips for new applications, including low crosstalk trapped-ion quantum computing[28,29].

In certain applications, the use of photonic devices with circular mode profiles is essential for efficient coupling with single-mode fibres. Through the proposed scheme, we have observed that highly circular modes can be readily achieved through high-quality multiscan processes. For instance, by applying four scans along the x-axis with a separation of 0.5 μm, we were able to easily create circular modes that precisely matched a single-mode fibre at a wavelength of 785 nm. Furthermore, we have achieved precise control over both the circularity and mode diameter. In our various application scenarios involving this technique, we successfully generated circular modes that matched single-mode fibres at wavelengths of 532 nm, 785 nm, and 1550 nm, with experimentally measured low coupling losses of around 0.19 - 0.39dB (Supplementary Table S2). The presented scheme offers a notable advantage in terms of low propagation loss as well. For example, in our experimental investigations, all devices exhibited measured propagation losses within the range of 0.15 - 0.5 dB/cm[3] at a wavelength of 785 nm.

## 3. Conclusion

We have presented several novel findings. Firstly, we conducted a comprehensive investigation on the impact of phase aberrations on refractive index modification in glass. Secondly, we proposed a new hypothesis model for refractive index modification in borosilicate glass, which qualitatively explains various experimental results for different types of focal intensity distributions. Thirdly, we introduce a novel index modification scheme utilizing a combination of lower and higher order spherical phases, which lead to an advantageous energy-relocated focal shape. The proposed index modification schemes combine the advantages of both non-thermal and thermal methods, enabling the rapid construction of structures with significantly higher refractive index contrast and high precision. The schemes offer distinct advantages in various application areas, including the development of high-precision, low-loss adiabatic mode converters, low-crosstalk integrated photonic chips for trapped-ion quantum computations, and any photonic chips requiring precise, non-symmetric, high-contrast refractive index modifications.

# 4. Methods

*Femto-second Laser Fabrication System*

The optical system and 3D spatial coordinators are included in supplementary information Figure S1. The ultrafast laser system used a regenerative amplified Yb:KGW laser (Light Conversion Pharos SP-06-1000-pp) with 1MHz repetition rate, 514 nm wavelength, 170 fs pulse duration. A Spatial Light Modulator (SLM, Hamamatsu Photonics X10468) was aligned and imaged by a 4-f lens system to the pupil of objective lens. The power of the laser beam was controlled with a motorized rotating half waveplate together with a Polarization Beam Splitter (PBS). The laser beam at objective lens focus was circularly polarized. The glass sample was fixed on a three-axis air bearing stage (Aerotech ABL10100L/ABL10100L/ANT95-3-V) to control the movements.

The experiments were conducted in borosilicate glass (Corning EAGLE 2000). For all the results in this paper, if not specified, the devices were fabricated using a 0.5NA objective lens (~93% transmission), depth of 120 μm from surface of glass sample, scanning speed of 8mm/s, pulse energy (measured at the objective pupil) was 80nJ for fabrication with $Z_{11} = -1$.

For all fabrication in this paper, the primary spherical aberration arising from the refractive index mismatch between immersion and sample was corrected, but not included in the description.

*Focal Intensity Simulation*

The Fourier optics model[26,30,31] was adopted to simulate 3D focal intensity distribution of ultrafast laser with different phase aberrations. The intensity was calculated by solving the Rayleigh-Sommerfeld diffraction integral from the pupil illumination.

*Microscopic Imaging*

After direct laser fabrication, photonic chips were polished by using a sequence of 30 μm, 9 μm, 3 μm and 1 μm polishing films. A layer of at least 150 μm glass were polished off both input and output facets of the glass. Supplementary Figure S2 includes a system diagram for microscopic images and loss measurements. A lab-built LED-illuminated widefield transmission microscope (microscope 1) was used to check device cross-sections after polishing. To image laser guiding mode profiles, single mode fibres were used to guide laser light into the input facet of a device. The fibre output was mounted and adjusted in a six-axis stage (three spatial axes plus three angle adjustment). Another lab-built microscope (microscope 2) was used to monitor and measure the distance between fibre tip and chip input facet. During the measurement of coupling losses, we brought the fibre tip to be close (< 1 μm) to the photonic chip to avoid input beam expansion. The output facet of a device was imaged by microscope 1 in order to capture guided laser mode profile.

*Refractive Index Profile Measurement*

Refractive index profiles were measured by a lab-built 3D tomographic microscope[32]. The imaging system recorded many intensity images of a photonic device at different illumination angles ranging between about -45° and 45°. The light source was a collimated blue LED (460 nm). A two-dimensional refractive index cross section of the photonic device was then reconstructed from the image stack using an error reduction algorithm based on gradient descent and simulated beam propagation.

**Acknowledgements**

This project was supported by UK Engineering and Physical Sciences Research Council (EPSRC) grants EP/X017931/1 and EP/V006797/1, as well as Austrian Science Fund (FWF) I3984-N36.

**Author contributions**

B.S. oversaw the project, proposed the hypothesis model, fabricated photonic waveguides and devices, obtained microscopic images and mode profiles, and performed various simulations. S.M. and A.J. developed the tomographic microscopy, measured, and analysed all refractive index profiles. A.J. wrote the phase algorithm of Nelder-Mead method and helped B.S. perform the phase optimization. P.S. provided the initial LabVIEW codes for SLM phase control. R.T. provided suggestions over thermal regime multiscan and contributed to the results analysis. B.S. and A.J. co-wrote the paper, with valuable inputs from P.S. M.B. contributed significantly to the paper writing, results analysis and supported the overall project. All authors discussed the results and reviewed the manuscript.

**Conflict of Interest**

The authors declare no competing interests.

# References

[1] W. Liu, M. Li, R. S. Guzzon, E. J. Norberg, J. S. Parker, M. Lu, L. A. Coldren, J. Yao, *Nat. Photonics* **2016**, *10*, 190.

[2] C. Sun, M. T. Wade, Y. Lee, J. S. Orcutt, L. Alloatti, M. S. Georgas, A. S. Waterman, J. M. Shainline, R. R. Avizienis, S. Lin, B. R. Moss, R. Kumar, F. Pavanello, A. H. Atabaki, H. M. Cook, A. J. Ou, J. C. Leu, Y. H. Chen, K. Asanović, R. J. Ram, M. A. Popović, V. M. Stojanović, *Nature* **2015**, *528*, 534.

[3] B. Sun, F. Morozko, P. S. Salter, S. Moser, Z. Pong, R. B. Patel, I. A. Walmsley, M. Wang, A. Hazan, N. Barré, A. Jesacher, J. Fells, C. He, A. Katiyi, Z. N. Tian, A. Karabchevsky, M. J. Booth, *Light Sci. Appl.* **2022**, *11*, 214.

[4] S. Gross, M. J. Withford, *Nanophotonics* **2015**, *4*, 332.

[5] G. Della Valle, R. Osellame, P. Laporta, G. Della Valle, R. Osellame, P. Laporta, G. Della Valle, *J. Opt. A Pure Appl. Opt.* **2009**, *11*, 13001.

[6] T. Meany, M. Gräfe, R. Heilmann, A. Perez-Leija, S. Gross, M. J. Steel, M. J. Withford, A. Szameit, *Laser Photonics Rev.* **2015**, *9*, 363.

[7] T. C. Ralph, J. B. Spring, B. J. Metcalf, P. C. Humphreys, W. S. Kolthammer, X.-M. Jin, M. Barbieri, A. Datta, N. Thomas-Peter, N. K. Langford, D. Kundys, J. C. Gates, B. J. Smith, P. G. R. Smith, I. A. Walmsley, *Nat. Photonics* **2013**, *7*, 514.

[8] B. J. Metcalf, J. B. Spring, P. C. Humphreys, N. Thomas-Peter, M. Barbieri, W. S. Kolthammer, X. M. Jin, N. K. Langford, D. Kundys, J. C. Gates, B. J. Smith, P. G. R. Smith, I. A. Walmsley, *Nat. Photonics* **2014**, *8*, 770.

[9] L. J. Maczewsky, M. Heinrich, M. Kremer, S. K. Ivanov, M. Ehrhardt, F. Martinez, Y. V. Kartashov, V. V. Konotop, L. Torner, D. Bauer, A. Szameit, *Science (80-. ).* **2020**, *370*, 701.

[10] L. J. Maczewsky, B. Höckendorf, M. Kremer, T. Biesenthal, M. Heinrich, A. Alvermann, H. Fehske, A. Szameit, *Nat. Mater.* **2020**, *19*, 855.

[11] S. Mukherjee, M. C. Rechtsman, *Science (80-. ).* **2020**, *368*, 856.

[12] A. J. Menssen, J. Guan, D. Felce, M. J. Booth, I. A. Walmsley, *Phys. Rev. Lett.* **2020**, *125*, 117401.

[13] J. Carolan, C. Harrold, C. Sparrow, E. Martín-López, N. J. Russell, J. W. Silverstone, P. J. Shadbolt, N. Matsuda, M. Oguma, M. Itoh, G. D. Marshall, M. G. Thompson, J. C. F. Matthews, T. Hashimoto, J. L. O'Brien, A. Laing, *Science (80-. ).* **2015**, *349*, 711.

[14] A. Politi, M. J. Cryan, J. G. Rarity, S. Yu, J. L. O'Brien, *Science (80-. ).* **2008**, *320*, 646.

[15] Y. Lei, H. Wang, L. Skuja, B. Kühn, B. Franz, Y. Svirko, P. G. Kazansky, *Laser Photonics Rev.* **2023**, *2200978*, 1.

[16] M. Beresna, P. G. Kazansky, *Adv. Opt. Photonics* **2014**, *339*, 293.

[17] M. Tillmann, B. Dakić, R. Heilmann, S. Nolte, A. Szameit, P. Walther, *Nat. Photonics* **2013**, *7*, 540.

[18] A. Crespi, R. Osellame, R. Ramponi, V. Giovannetti, R. Fazio, L. Sansoni, F. De Nicola, F. Sciarrino, P. Mataloni, *Nat. Photonics* **2013**, *7*, 322.

[19] L. Sansoni, F. Sciarrino, G. Vallone, P. Mataloni, A. Crespi, R. Ramponi, R. Osellame, *Phys. Rev. Lett.* **2012**, *108*, 1.

[20] C. Antón, J. C. Loredo, G. Coppola, H. Ollivier, N. Viggianiello, A. Harouri, N. Somaschi, A. Crespi, I. Sagnes, A. Lemaître, L. Lanco, R. Osellame, F. Sciarrino, P. Senellart, *Optica* **2019**, *6*, 1471.

[21] X. Y. Xu, X. W. Wang, D. Y. Chen, C. M. Smith, X. M. Jin, *Nat. Photonics* **2021**, *15*, 703.

[22] A. Crespi, F. V. Pepe, P. Facchi, F. Sciarrino, P. Mataloni, H. Nakazato, S. Pascazio, R. Osellame, *Phys. Rev. Lett.* **2019**, *122*, 130401.

[23] Y. Nasu, M. Kohtoku, Y. Hibino, *Opt. Lett.* **2005**, *30*, 723.

[24] R. R. Thomson, T. A. Birks, S. G. Leon-Saval, A. K. Kar, J. Bland-Hawthorn, *Opt. Express* **2011**, *19*, 5698.

[25] D. G. MacLachlan, R. J. Harris, D. Choudhury, R. D. Simmonds, P. S. Salter, M. J. Booth, J. R. Allington-Smith, R. R. Thomson, *Opt. Lett.* **2016**, *41*, 76.

[26] M. Born, E. Wolf, *Principles of Optics*, Cambridge University, **2010**.

[27] P. S. Salter, M. J. Booth, *Light Sci. Appl.* **2019**, *8*, 1.

[28] A. Sotirova, B. Sun, J. Leppard, A. Wang, M. Wang, A. V. Brennan, F. Pokorny, D. Nadlinger, C. He, P. S. Salter, M. J. Booth, C. Ballance, *Prep.* **2023**.

[29] M. L. Day, K. Choonee, Z. Chaboyer, S. Gross, M. J. Withford, A. G. Sinclair, G. D. Marshall, *Quantum Sci. Technol.* **2021**, *6*, 024007.

[30] B. Sun, P. S. Salter, M. J. Booth, *J. Opt. Soc. Am. A* **2014**, *31*, 765.

[31] B. Sun, P. S. Salter, M. J. Booth, *J. Opt. Soc. Am. B* **2015**, *32*, 1272.

[32] N. Barré, R. Shivaraman, L. Ackermann, S. Moser, M. Schmidt, P. Salter, M. Booth, A. Jesacher, *Opt. Express* **2021**, *29*, 35414.

# Supplementary Information

## High Speed Precise Refractive Index Modification for Photonic Chips through Phase Aberrated Pulsed Lasers

Bangshan Sun[1*], Simon Moser[2], Alexander Jesacher[2*], Patrick S. Salter[1], Robert R. Thomson[3], Martin J. Booth[1*]

[1]Department of Engineering Science, University of Oxford, Oxford OX1 3PJ, United Kingdom
[2]Institute of Biomedical Physics, Medical University of Innsbruck, Müllerstraße 44, 6020 Innsbruck, Austria
[3]Institute of Photonics and Quantum Sciences, Heriot-Watt University, Edinburgh EH14 4AS, UK

*Corresponding to: Bangshan Sun (bangshan.sun@eng.ox.ac.uk), or Alexander Jesacher (alexander.jesacher@i-med.ac.at), or Martin J. Booth (martin.booth@eng.ox.ac.uk)

# Note1. Accumulated intensity profiles calculation

In Figure 1 (c) and Figure 3 (b) of main text, we present the accumulated intensity profiles along x and z axes, represented as $I(x, z)$ here. The calculation of these accumulated intensity profiles is as follows. Please refer Supplementary Figure S1 for the spatial axes.

$$I(x, z) = \int_{-d}^{d} I(x, y, z) \, dy$$

Here, we use the distance between two first minimum intensity in airy pattern to define integration distance of d. Airy pattern[1] (or called Airy disk) is used for an accurate description of the optimal focused spot of light that a perfect lens with a circular aperture can make. In lateral focal plane (x-y plane), the physical distance between the two first minimum intensity beside the peak value can be expressed by,

$$d = \frac{1.22\lambda}{NA}$$

where NA represents the numerical aperture of objective lens.

## Note2. Device loss measurements

To analyse the propagation loss and coupling loss of the waveguides, a cut-back approach was adopted. A power meter was placed at camera position of microscope to measure laser powers: 1) directly at fibre tip without waveguide sample ($P_{fiber}$); 2) at waveguide output facet after full-length waveguide ($P_{output\_long}$); 3) at waveguide output facet after cutting the waveguide to a shorter length ($P_{output\_short}$), where the lengths of waveguide ($Length_{long}$, $Length_{short}$) were measured in centimetres. The waveguide coupling efficiency ($E_c$) is defined as proportion of laser light coupled from the waveguide to the other waveguide/fibre. The propagation efficiency per centimetre ($E_p$) is defined as proportion of laser light transmitted after one centimetre of waveguide. $E_c$ and $E_p$ can be resolved by solving below two equations with the measurements of longer and shorter waveguides,

$$P_{output\_long} = P_{fiber} \cdot E_c \cdot (E_p)^{Length_{long}}$$

$$P_{output\_short} = P_{fiber} \cdot E_c \cdot (E_p)^{Length_{short}}$$

Applying the relation between Efficiency ($E$) and Loss (in terms of dB), $Loss = 10 \cdot \log_{10} E$, the coupling loss ($Loss_c$) and propagation loss ($Loss_p$) can also be directly calculated.

# Table S1. Zernike polynomials

We investigated individually common Zernike modes in Noll's sequential indices[1] with an RMS amplitude of -1 rad, including Z5 – Z10, as well as primary and higher order spherical aberrations. The details of these Zernike phase aberrations are listed as follows.

Table S1. The Zernike polynomials k = 5 to 10, and spherical aberrations

| k | n | m | $Z_k(\rho, \theta)$ | Aberration Term |
|---|---|---|---|---|
| 5 | 2 | -2 | $\sqrt{6}\rho^2 \sin(2\theta)$ | Astigmatism |
| 6 | 2 | 2 | $\sqrt{6}\rho^2 \cos(2\theta)$ | Astigmatism |
| 7 | 3 | -1 | $2\sqrt{2}(3\rho^3 - 2\rho)\sin(\theta)$ | Coma |
| 8 | 3 | 1 | $2\sqrt{2}(3\rho^3 - 2\rho)\cos(\theta)$ | Coma |
| 9 | 3 | -3 | $2\sqrt{2}\rho^3 \sin(3\theta)$ | Trefoil |
| 10 | 3 | 3 | $2\sqrt{2}\rho^3 \cos(3\theta)$ | Trefoil |
| 11 | 4 | 0 | $\sqrt{5}(6\rho^4 - 6\rho^2 + 1)$ | Primary Spherical |
| 22 | 6 | 0 | $\sqrt{7}(20\rho^6 - 30\rho^4 + 12\rho^2 - 1)$ | 2nd order Spherical |
| 37 | 8 | 0 | $3 \cdot (70\rho^8 - 140\rho^6 + 90\rho^4 - 20\rho^2 + 1)$ | 3rd order Spherical |
| 56 | 10 | 0 | $\sqrt{11}(252\rho^{10} - 630\rho^8 + 560\rho^6 - 210\rho^4 + 30\rho^2 - 1)$ | 4th order Spherical |

## Table S2. Typical losses of photonic devices

By employing the presented new fabrication scheme, circular cross-sections can be readily created to align with single-mode fibres. These versatile circular shapes can be tailored to different diameters, making them suitable for applications across various wavelengths[2]. Table S2 presents the typical losses for these devices, with measurements conducted as outlined in Supplementary Note 2. Notably, shorter wavelengths like 532nm exhibit higher coupling loss, likely attributed to the challenges in alignment owing to their smaller mode size. Additionally, higher propagation loss can be attributed to increased optical glass absorption and higher sensitivity to defects at shorter wavelengths. All these devices were fabricated with scanning speed of 4mm/s. Immersion oil was applied to increase coupling.

Table 2. The losses measured for devices fabricated for varying wavelengths. Mode field diameter is the diameter at which the optical power is reduced to $1/e^2$ from its peak level.

| Devices | Wavelengths (nm) | Mode diameter | Coupling loss (dB) | Propagation loss (dB/cm) |
|---------|------------------|---------------|--------------------|--------------------------|
| 1 | 1550 | 9.5 μm | 0.19 dB | 0.15 dB/cm |
| 2 | 785 | 5.2 μm | 0.22 dB | 0.14 dB/cm |
| 3 | 532 | 3.6 μm | 0.39 dB | 0.48 dB/cm |

Figure S1. Sketch of ultrafast laser fabrication system and spatial axes

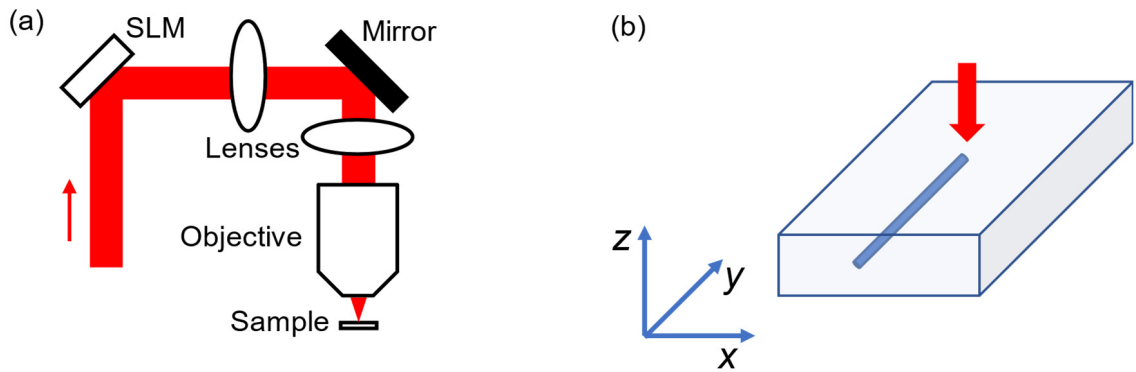

Figure S1. (a) A sketch of optical system. SLM: spatial light modulator. (b) Sketch to illustrate the 3D coordinators in the relative to the sample. Red arrow marks the laser propagation direction.

Figure S2. Photonic chip characterization system

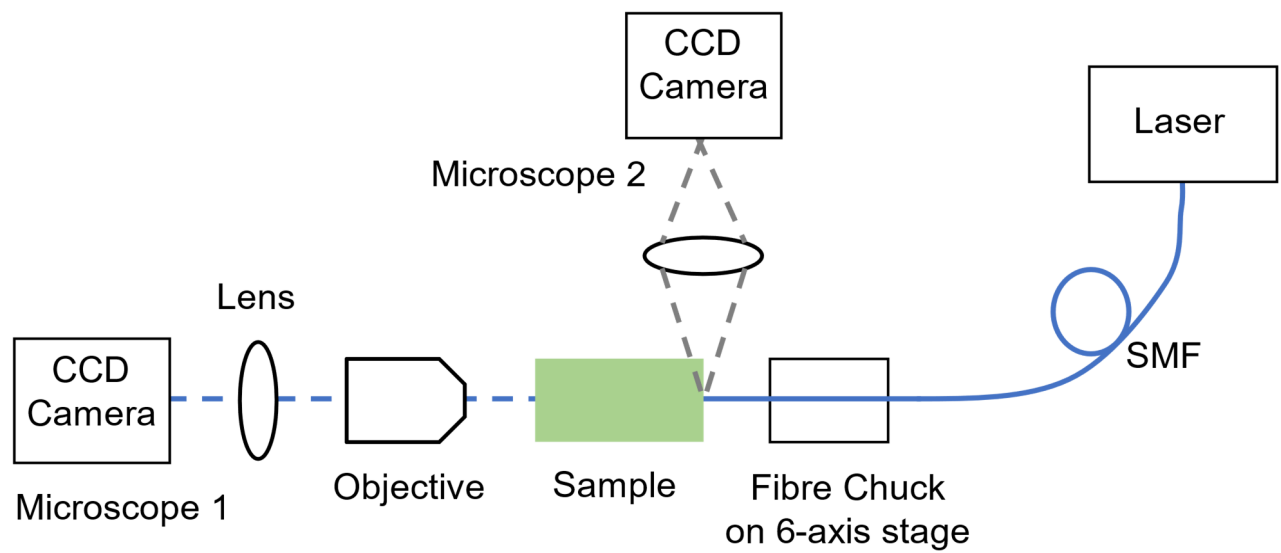

Figure S2. The system to image device facets, obtain laser mode profiles, and measure photonic device losses. There were two lab-built microscopes: microscope 1 is used to obtain device facet images and mode profiles; microscope 2 was used for live monitoring and measurement of the distance between input fibre tip and device facet.

Figure S3. Enlarged refractive index profiles for main text Figure 3

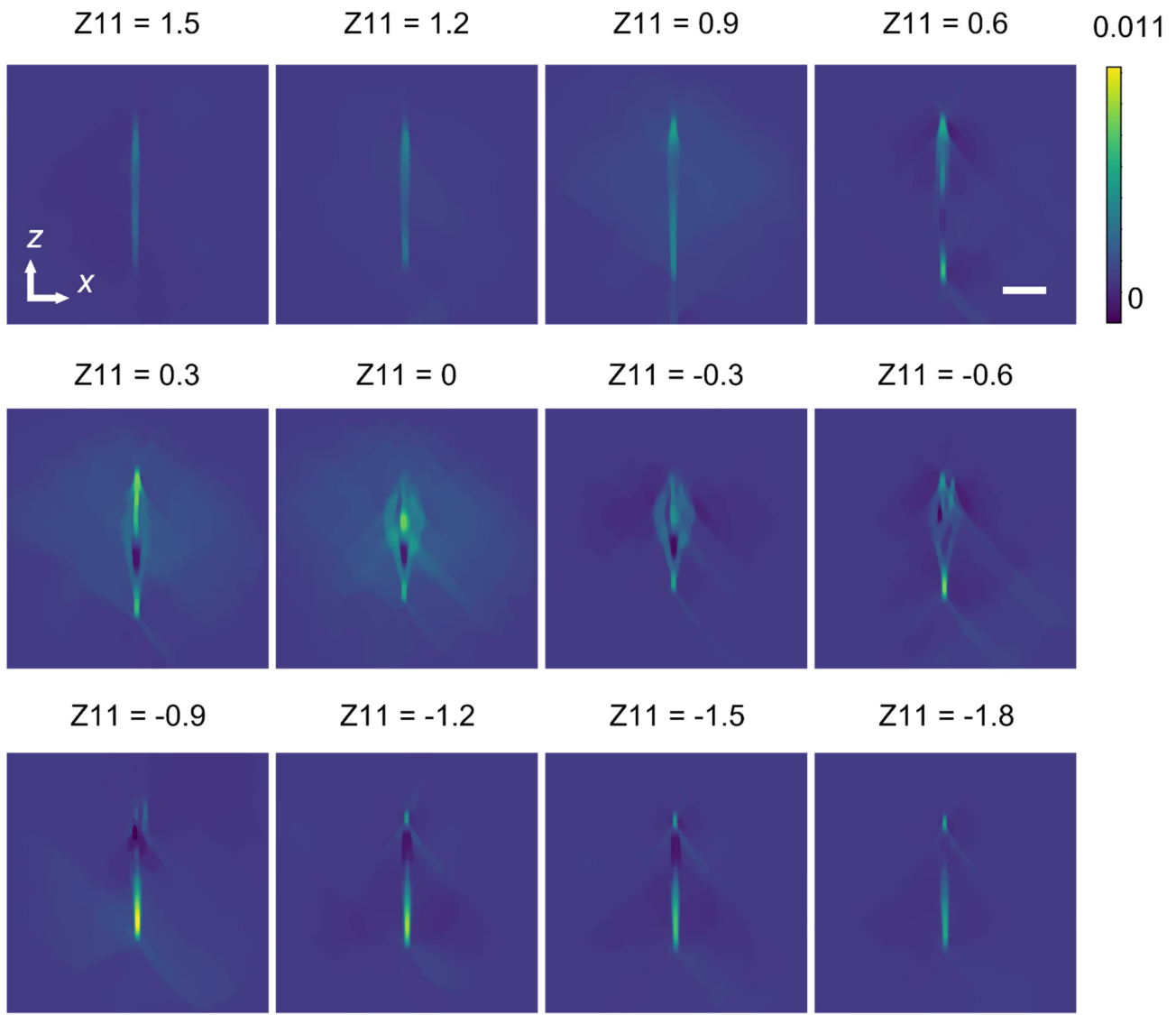

Figure S3. Enlarged refractive index profiles to reveal more details for main text Figure 3. 3D tomographic microscopic refractive index images of waveguide cross-sections fabricated with varying Z11 phase applied. Scale bars: 5 μm.

Figure S4. Enlarged refractive index profiles for main text Figure 4

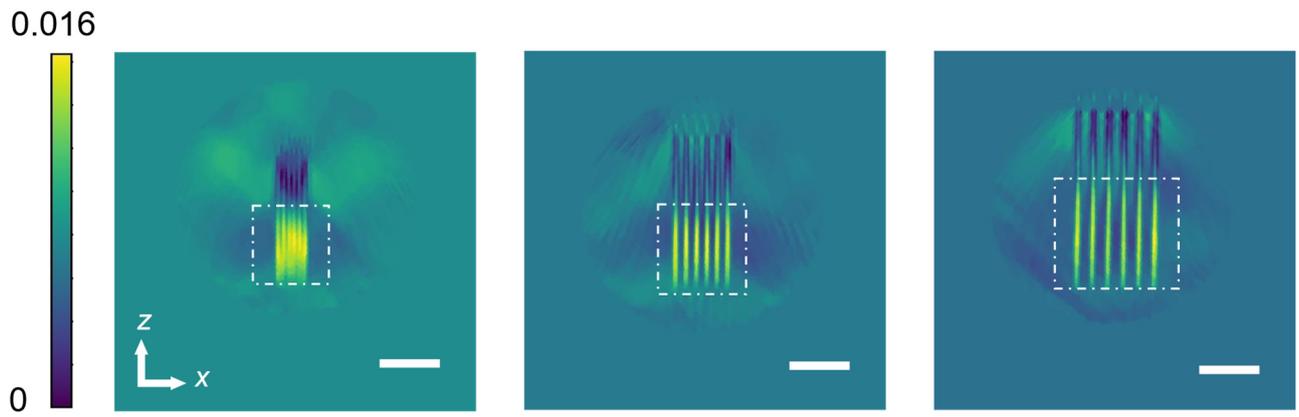

Figure S4. Enlarged refractive index profiles to reveal more details for main text Figure 4. Multiscan with varying scan spacings. Scale bars: 5 μm.

Figure S5. Enlarged refractive index profiles for main text Figure 5

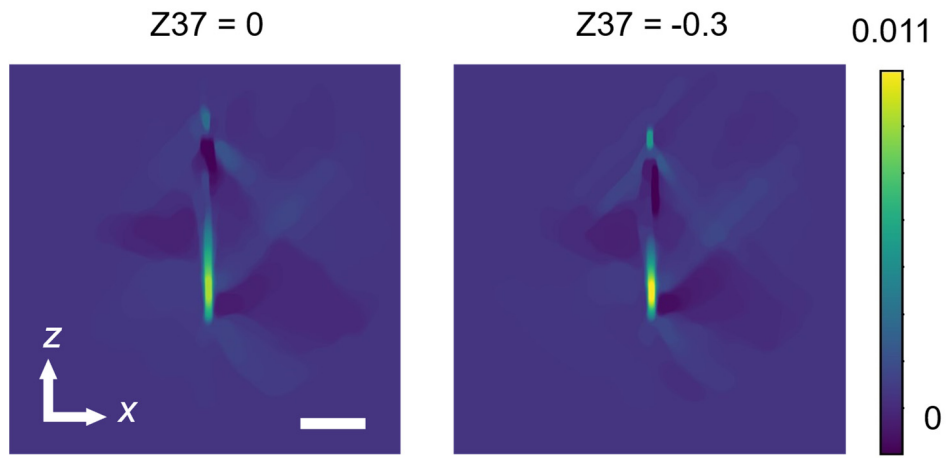

Figure S5. Enlarged refractive index profiles to reveal more details for main text Figure 5. Comparing the case without Z37 to the case with an additional amplitude of Z37 = -0.3. Scale bars: 5 μm.

**References**

[1] M. Born, E. Wolf, *Principles of Optics*, Cambridge University, **2010**.

[2] B. Sun, F. Morozko, P. S. Salter, S. Moser, Z. Pong, R. B. Patel, I. A. Walmsley, M. Wang, A. Hazan, N. Barré, A. Jesacher, J. Fells, C. He, A. Katiyi, Z. N. Tian, A. Karabchevsky, M. J. Booth, *Light Sci. Appl.* **2022**, *11*, 214.